\documentclass{emulateapj}
\usepackage{natbib}

\newcommand{\src}{KS~1947+300}
\newcommand{\eps}{{\rm ergs\,s^{-1}}}
\newcommand{\epcs}{{\rm ergs\,cm^{-2}\,s^{-1}}}
\newcommand{\xte}{{\it RXTE}}
\newcommand{\cts}{{\rm count\,s^{-1}}}

\newcommand{\delf}{$1.8\ \mu$Hz}		
\newcommand{\delffrac}{$3.7\times10^{-5}$}
\newcommand{\tglitch}{MJD~51956.41}
\newcommand{\porb}{$40.415\pm0.010$~d}
\newcommand{\ecc}{0.033\pm0.013}
\newcommand{\axsini}{$137 \pm 3$~lt-sec}
\newcommand{\fxm}{$1.71 \pm 0.04\ M_\sun$}
\newcommand{\maxfdot}{$2.3\times10^{-11}\ {\rm Hz\,s^{-1}}$} 

\slugcomment{Accepted by ApJ}	

\shortauthors{Galloway, Morgan \& Levine}
\shorttitle{A Frequency Glitch in an Accreting Pulsar}

\begin{document}

\title{A Frequency Glitch in an Accreting Pulsar}

\author{D.~K. Galloway, E.~H. Morgan, A. M. Levine}
\affil{Center for Space Research, MIT,
    37-626b, 77 Massachusetts Avenue, Cambridge, MA 02139-4307}
\email{duncan, ehm, aml@space.mit.edu}

\begin{abstract}

Frequency glitches have been observed so far only in radio pulsars and
anomalous X-ray pulsars.  Here we present evidence of a glitch in a
neutron star accreting from a Be companion.
The transient KS~1947+300 reappeared in 2000 October as a moderately
strong X-ray source that
exhibited 18.7~s pulsations, leading to an identification with the BATSE
source GRO~J1948+32, last detected in 1994.  We have analyzed Rossi X-ray
Timing Explorer observations taken during the 2000--01 outburst, as well
as additional observations taken during a smaller outburst in July 2002.
Orbital Doppler shifts are apparent in the temporal variation of the pulse
frequency.  A joint fit of the \xte\/ data together with BATSE
measurements from an outburst in 1994 yields the orbital period $P_{orb}
=$ \porb, the projected orbital radius $a_{\rm x} \sin i = 137 \pm 3$
lt-s, and the eccentricity $e = 0.033 \pm 0.013$.  This degree of
eccentricity is unexpectedly low for such a wide orbit.  Pulse timing
results also show that the intrinsic pulse frequency increased from 53.30
to 53.47 mHz at a rate approximately proportional to the X-ray flux.  This
is about the degree of spin up expected from the accretion torques that
must be present when the X-ray luminosity reaches $\sim10^{38}$ ergs
s$^{-1}$.  On one occasion during the 2000--01 outburst, the pulse
frequency increased by $\sim 1.8 \times 10^{-6}$ Hz in $\lesssim 10$~hr
over and above the mean trend seen around that time, without any indication
of a correspondingly large increase in X-ray flux.
The fractional change in frequency of $3.7\times10^{-5}$ during this event
is significantly larger than the values observed in the glitches in radio
pulsars and anomalous X-ray pulsars. We discuss other similarities and
differences between these events.

\end{abstract}


\section{INTRODUCTION}

Abrupt changes (``glitches'') in the spin frequency of rotating
neutron stars have been observed to date in radio pulsars
\cite[e.g.][]{glitch03} and anomalous X-ray pulsars \cite[AXPs;
e.g.][]{kaspi03,kaspi03b}.  These events appear as sudden increases in
the observed pulse frequency, superimposed on more-or-less steady
spin-down arising from magneto-dipole braking, and are attributed to a
sudden transfer of angular momentum between the superfluid of the
interior and the crust \cite[]{ruderman76,ai75}.  Between glitches the
interior superfluid and the crust are relatively loosely coupled, and
the braking torque acts primarily on the crust, so that just prior to
a glitch the crust is spinning slower than the interior superfluid.

To date there have been no reports of glitches in accreting pulsars.
Here we present the results of a pulse timing analysis of data
obtained from {\it Rossi X-ray Timing Explorer}\/ ({\it RXTE})
observations of the accreting pulsar \src, from which we derive
orbital parameters as well as the intrinsic spin frequency history of
the neutron star.  The latter includes a short time interval during
which the pulse frequency increased at an unusually high rate, that
may be the first evidence for a frequency glitch similar to
those observed in radio pulsars and AXPs.

The cosmic X-ray source \src\ ($l^{II}=66\fdg09$, $b^{II}=+2\fdg10$)
was first detected on 1989 June 8 with the TTM coded mask imaging
spectrometer mounted on {\it Mir's} Kvant module
\citep{skin89,borozdin90}.  The optical counterpart is a $V=14.2$ B0Ve
star with moderate reddening that indicates the distance to the system
is $\approx10$~kpc \cite[]{neg02}.  In 1994 the Burst and Transient
Source Experiment (BATSE) aboard the {\it Compton Gamma Ray
Observatory (CGRO)} detected 18.7~s pulsations from an X-ray source 
within a few degrees of \src.  From the BATSE data, \cite{chak95}
found that the apparent pulse period of GRO~J1948+32, as the pulsating
source was designated, varied systematically over the 33~d outburst,
indicating that the source was in a binary system with an orbital
period between 35 and 44 days.  In 2000 November, weak emission from
\src\/ was detected with the All-Sky Monitor (ASM) aboard \xte.
A search for periodicities in the $\sim 5$ yr ensemble of ASM measurements
of \src\/ accumulated up until that time yielded evidence of a
$41.7\pm0.1$~day periodicity \citep{levine00}.  Soon afterwards,
pulsations with a period of 18.7~s were found in follow-up observations
with the Proportional Counter Array (PCA) on {\it RXTE}, which showed that
\src\ and GRO~J1948+32 are one and the same \citep{swank00b}.  A series of
{\it RXTE}\/ target-of-opportunity (TOO) observations were then conducted
which, together with the measurements by the ASM, gave excellent coverage
over a long and strong outburst.

\section{OBSERVATIONS AND ANALYSIS}
\label{fluxobs}

We made observations of \src\ with all three instruments aboard \xte: the
ASM, PCA and the High-Energy X-ray Timing Experiment (HEXTE).
The ASM views much of the sky every few hours \cite[]{asm96}.
The instrument consists of three Scanning Shadow Cameras (SSCs)
mounted on a rotating platform, 
and carries out sequences of 90~s
observations
\centerline{\epsfxsize=8.5cm\epsfbox{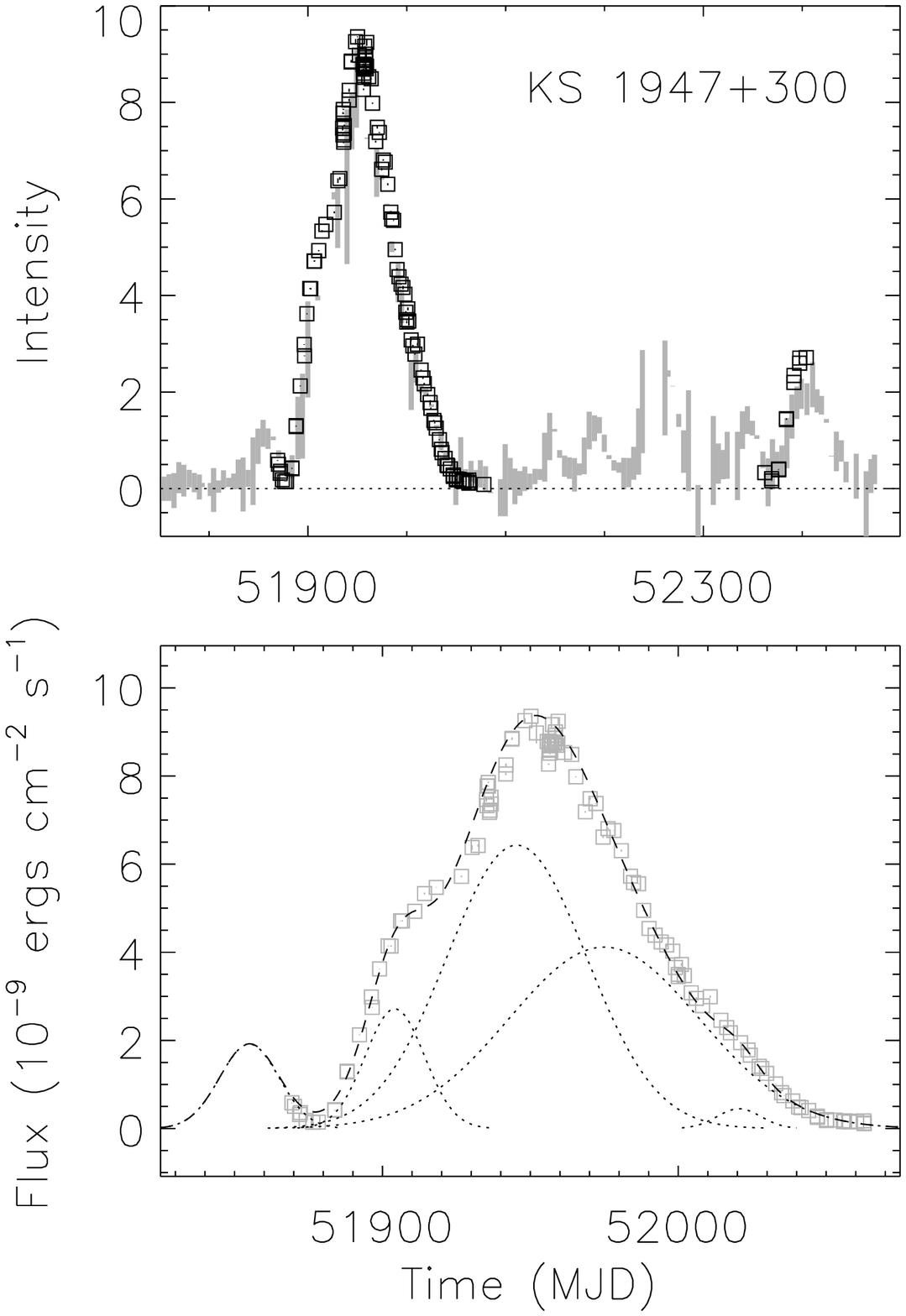}}
 \figcaption[f1.eps]{Intensity of \src\ throughout the 2000--01 outburst.
({\it upper panel}) The gray
markers show the ASM count rate (1.5--12 keV) $1\sigma$ confidence
regions (with scale on the left in ASM counts s$^{-1}$), from 1 day
averages of 90~s dwells, smoothed over a 5-d window (excluding those
with estimated 1$\sigma$ errors greater than $0.7\ \cts$).  The open
squares show the integrated 2--60~keV source flux (with scale on the
left in units of $10^{-9}\ \epcs$) as measured from best-fit models of
PCA and HEXTE spectra.  MJD 52000 corresponds to 2001 April 1.  ({\it
lower panel}) The flux measurements are plotted as squares on an
expanded time axis, along with the individual Gaussian components
(dotted lines) used to model the flux evolution, and the total model
(dashed line; see \S\ref{fmodel}).
 \label{asmlc} }
\bigskip
\noindent
called dwells, in between which the platform holding the
SSCs is rotated by $6\arcdeg$. The data from
each SSC are analyzed to obtain estimates of the intensities of all the
sources listed in the ASM analysis catalog within the field of view.  The
intensities are expressed as the count rate expected from a source at the
center of the field of view of SSC 1 in 1996 March\footnote{The Crab
Nebula appears in the ASM as a 75 $\cts$ source (1.5-12 keV) and has a
corresponding energy flux in that energy band of $\sim2.7 \times 10^{-8}\
\epcs$.}

The ASM 1.5--12~keV intensity measurements of \src\ from 2000 July
through 2002 April are shown in Fig. \ref{asmlc}.  The source became
active in 2000 October and rose to an intensity of approximately
1~$\cts$ around 2000 November 7 (MJD~51855).
Following this initial peak the intensity
declined somewhat before beginning a second, more substantial increase
which peaked around MJD~51950 at $\sim9~\cts$ (approximately
120~mCrab).  Following the outburst maximum, the source declined to
$< 1~\cts$, but subsequently exhibited a number of small outbursts peaking at
1--3~$\cts$ and lasting for 10--50~d.  The source was
not detected by the ASM prior to MJD 51800; the intensity was
$\lesssim 0.2\ \cts$.  Relatively few measurements were taken around
MJD 51960, and again around MJD 52290, when the Sun was relatively
near its closest angular distance of $\sim 50\arcdeg$ from the source.
In addition, the measurements taken close to those times were more
strongly affected by systematic problems than measurements at other
times.

In 2000 November, shortly after the outburst was noticed in the ASM
data, a series of observations with the 
PCA and HEXTE aboard \xte\
was initiated.  The PCA consists of 5 Proportional Counter Units
(PCUs) sensitive to X-ray photons in the range 2.5--90~keV, each with a
collecting area of $\sim 1400\ {\rm cm^2}$ and a $1\arcdeg$
field of view \cite[]{xte96}.  Photon arrival times are measured to
$\approx1\ \mu{\rm s}$.  Spectra are accumulated in up to 256 energy
channels.
The number of PCUs operating during the observations varied between 2 and 5.
The HEXTE comprises two clusters, each of which contains 4 scintillation
detectors sensitive to photons in the range 15--250~keV, collimated to
view a common $1\arcdeg$ field \cite[]{hexte96}.  The detectors in the two
clusters provide a total collecting area of $1600\ {\rm cm^2}$.

The first PCA/HEXTE observation of \src\ was performed on 2000
November 21 (MJD 51869).  Subsequently, 1--6~ks long PCA/HEXTE
observations were made every 1--2 days until the end of the main
outburst around 2001 June 18 (MJD 52078).  Near the peak of the main
outburst two long observations were made (for proposal ID 50068, PI:
W. Heindl) that yielded 51 and 98~ks of exposure, respectively.  In
order to obtain a longer baseline for determination of the orbital
period, an additional series of observations began on 2002 March 29
(MJD~52362) and continued until 2002 May 17 (MJD~52411). In all, 132
observations were made with a total observing time of 371~ks.

Spectral analysis of the PCA and HEXTE data was undertaken using {\sc
lheasoft} release 5.1 (2001 June 26).  The data were first screened to
ensure that the center of the fields of view were within $0.02\arcdeg$ of
the direction of \src\/ and that the limb of the Earth was more than
$10\arcdeg$ from the source direction.
Mean energy, i.e., pulse height, spectra were extracted for each
observation from standard observing modes (``Standard-2'', with 129
channels between 2--60~keV for PCA, and ``Archive'', with 64 channels
between 15--250~keV for HEXTE).  PCA spectra were accumulated separately
for each PCU.
Instrument response matrices were generated for each PCU and each
observation using {\sc pcarsp}.
We estimated background count rates in the PCA using either faint or
bright source models (depending upon whether the count rate exceeded
$\approx40\ {\rm counts\,s^{-1}\,PCU^{-1}}$) developed for gain epoch 5
(from 2000 May 13 onwards) with {\sc pcabackest} version 2.1e.
We measured fluxes by integrating model fits to PCA and HEXTE spectra (see
\S\ref{specprof}).

Pulse timing analyses were performed using software written by the
\xte\ team at MIT. For timing analysis, the 3-12 keV energy range was
found to provide nearly optimal signal-to-noise ratios during the
faintest observations in which we could detect pulsations.  Thus we
selected, from ``GoodXenon'' mode data, only those events within the
energy range 3-12 keV detected in the top layer of any PCU, adjusted
the times of the events according to the projected distance of the
spacecraft from the Solar System barycenter, and binned the selected
events into 0.25 s time bins.

\section{SPECTRUM}
\label{specprof}

We fitted various continuum models to PCA spectra in the range 2--25~keV
and HEXTE spectra in the range 15--80~keV using {\sc xspec} version 11.0.
Above a flux of $\approx8\times10^{-10}\ \epcs$ (2--60~keV) the best fit
was obtained with a combination of Comptonization \cite[`{\tt compTT}' in
{\sc xspec}]{tit94} and blackbody components, along with a Gaussian
representing fluorescent Fe K$\alpha$ emission centered around 6.5~keV.
Other models typically used for pulsar spectra, including
cutoff or broken power laws, or a power law with a blackbody component,
generally gave worse fits. At high count rates, systematic errors in the
PCA response are known to contribute to deviations from the model; for
example, the reduced-$\chi^2$ ($\chi^2_\nu$) for the Comptonization model
fit to the highest flux spectrum was 3.41 (i.e. $\chi^2=511.7$ for 150
degrees of freedom). Adopting a systematic error of 1.7\% (2\% is often
used for the PCA) was sufficient to reduce the fit statistic to 1.01. For
comparison, a fit with this assumed systematic error and the same Gaussian
component combined with a power law with a high-energy exponential cutoff
gave $\chi^2_\nu=1.52$, while a combination of blackbody and powerlaw
components gave 5.29.

The PCA spectrum was significantly softer near the peak of the outburst 
compared to spectra obtained during the rise and decay 
(Fig. \ref{twospec}).
This was primarily due to a decreased contribution from the $T_{\rm
bb}=3$--4~keV blackbody component at higher fluxes, rather than a
systematic variation in $T_{\rm bb}$ (for example).
While $T_{\rm bb}$ is unusually high for a neutron star,
examination of the unfolded spectra suggests that the requirement for the
additional component may be due to intrinsic curvature of the
approximately power law part of the Comptonized spectra.  A similarly good
fit was obtained with a very broad ($\sigma\approx 10$~keV) Gaussian
component in place of the blackbody, and we note that the integrated flux
is not sensitive to the nature of the additional component.  Regardless of
whether a blackbody or Gaussian component is used, the additional
component is strongly required for the best fit.

The fitted column densities of neutral (unionized) matter measured
over the outburst were consistent with the line-of-sight value
estimated from H{\sc i}/dust surveys, i.e., $\approx 10^{22}\ {\rm atoms\
cm^{-2}}$ \cite[]{dl90,dust98}. Thus, to obtain flux estimates we fixed
$n_{\rm H}$ at this value and integrated the best-fit spectral models
comprising Comptonization and blackbody components over the 2-60 keV band.
The results are shown in Fig. \ref{asmlc}.
At the peak of the outburst in 2001 February the 2--60~keV flux was $\sim
9 \times 10^{-9}\ \epcs$.  The X-ray luminosity was therefore $\sim 1.1
\times 10^{38}\ \eps$ for a source distance of 10~kpc.

\section{PULSE TIMING ANALYSIS}

\subsection{Frequency measurements}
\label{freq}

To measure the long term frequency evolution of the pulsar we grouped
together those \xte\/ observations
\centerline{\epsfxsize=8.5cm\epsfbox{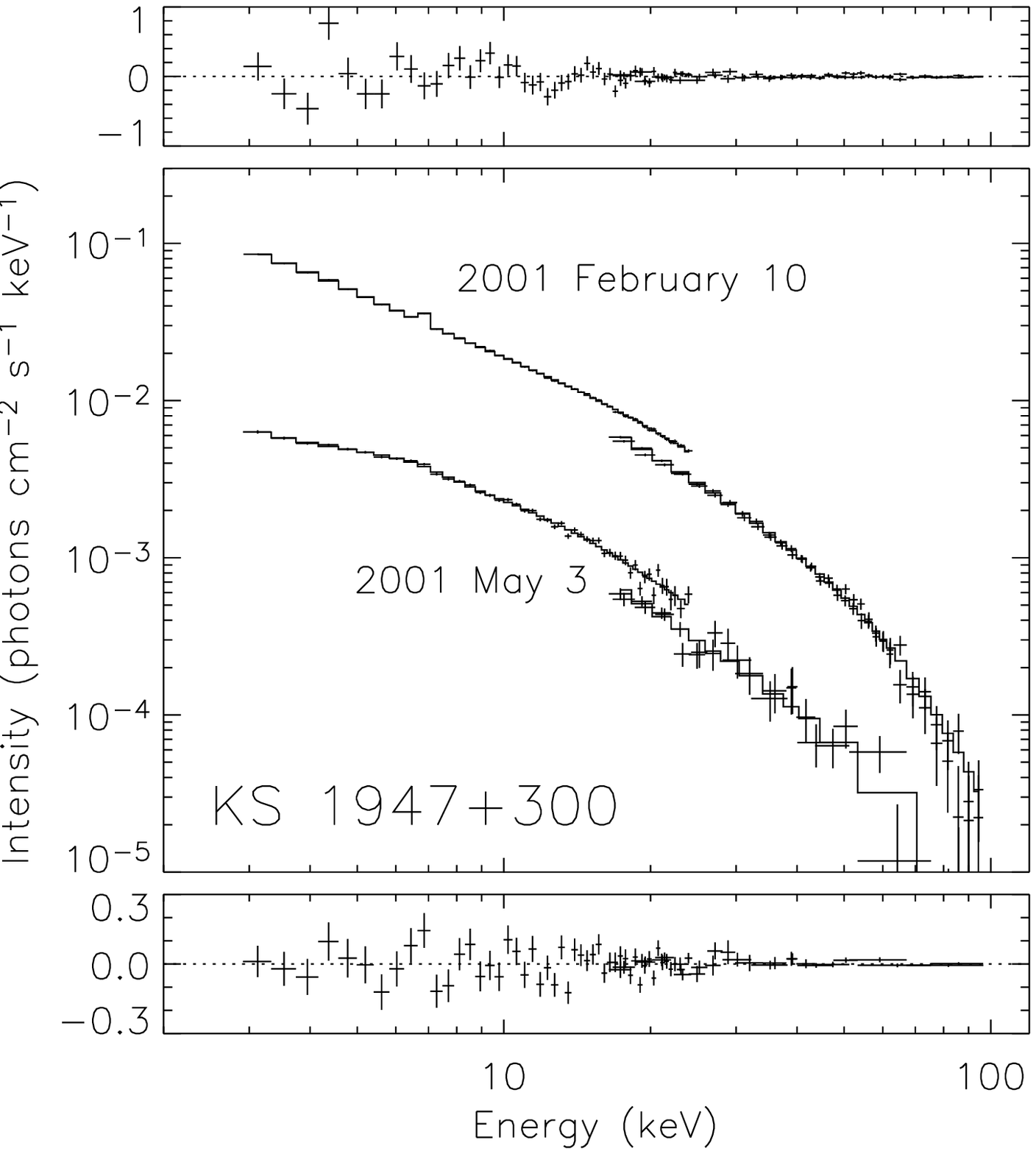}}
 \figcaption[twospec.eps]{Representative photon spectra of \src\ from the
peak (2001 February 10, or MJD~51950) and late in the decay (2001 May 3,
MJD~52032) of the 2000--01 outburst. The effect of the instrumental
responses have been removed (``unfolded''). Data from PCU \#2 are shown,
along with data from HEXTE clusters 1 and 2. For the 2001 May 3
observation, we rebinned the HEXTE spectra by a factor of 4. Error bars
indicate the $1\sigma$ uncertainties.
The top and bottom panels show the residuals to the model fit (units of
normalized counts$\,{\rm s^{-1}\ keV^{-1}}$) for the 2001 February 10 and
2001 May 3 spectra, respectively.
 \label{twospec} }
\bigskip
\noindent
that began within 8~hr of each other and would form a 
combined data set spanning less than 24~hr.
The frequency was then estimated for each set of observations by
fitting the 0.25-s time series with the function
\begin{equation}
F(t) = a_0 + a_1 \sin(\omega t) + a_2 \cos(\omega t) + a_3 \sin (2 \omega t)
 + a_4 \cos(2 \omega t)
\end{equation}
over a grid of trial frequencies where the coefficients $a_i$ were
chosen to obtain the best fit for each frequency $\omega$.  The
accuracy of each frequency measurement was estimated by the range of
frequencies where $\chi^2\leq\chi^2_{min}+10$ (equivalent to
$\approx3\sigma$), where $\chi^2_{min}$ was the value of the goodness
of fit statistic for the best fit.

We also used pulse frequencies determined from observations performed
in 1994 with the BATSE instrument on the {\it Compton Gamma Ray
Observatory}.  The frequencies were obtained in the same way as those
reported by \cite{chak95} except for being determined from 1-day
rather than 2-day data sets.  Data for the photon energy range 20--60
keV were used in this analysis.

The measured frequencies are shown in Fig. \ref{period}.  In the
$\approx6$ year interval between the 1994 and 2000 outbursts the neutron
star rotational frequency decreased from 53.47 to
53.30~mHz, corresponding to a mean spin-down rate $\dot{f}
\approx -8.2\times10^{-13}\ {\rm Hz\,s^{-1}}$.  During the 2000--2001
outburst, there was a substantial overall increase in pulse frequency
likely caused by accretion torques on the neutron star.  An
approximately sinusoidal periodic
\centerline{\epsfxsize=8.5cm\epsfbox{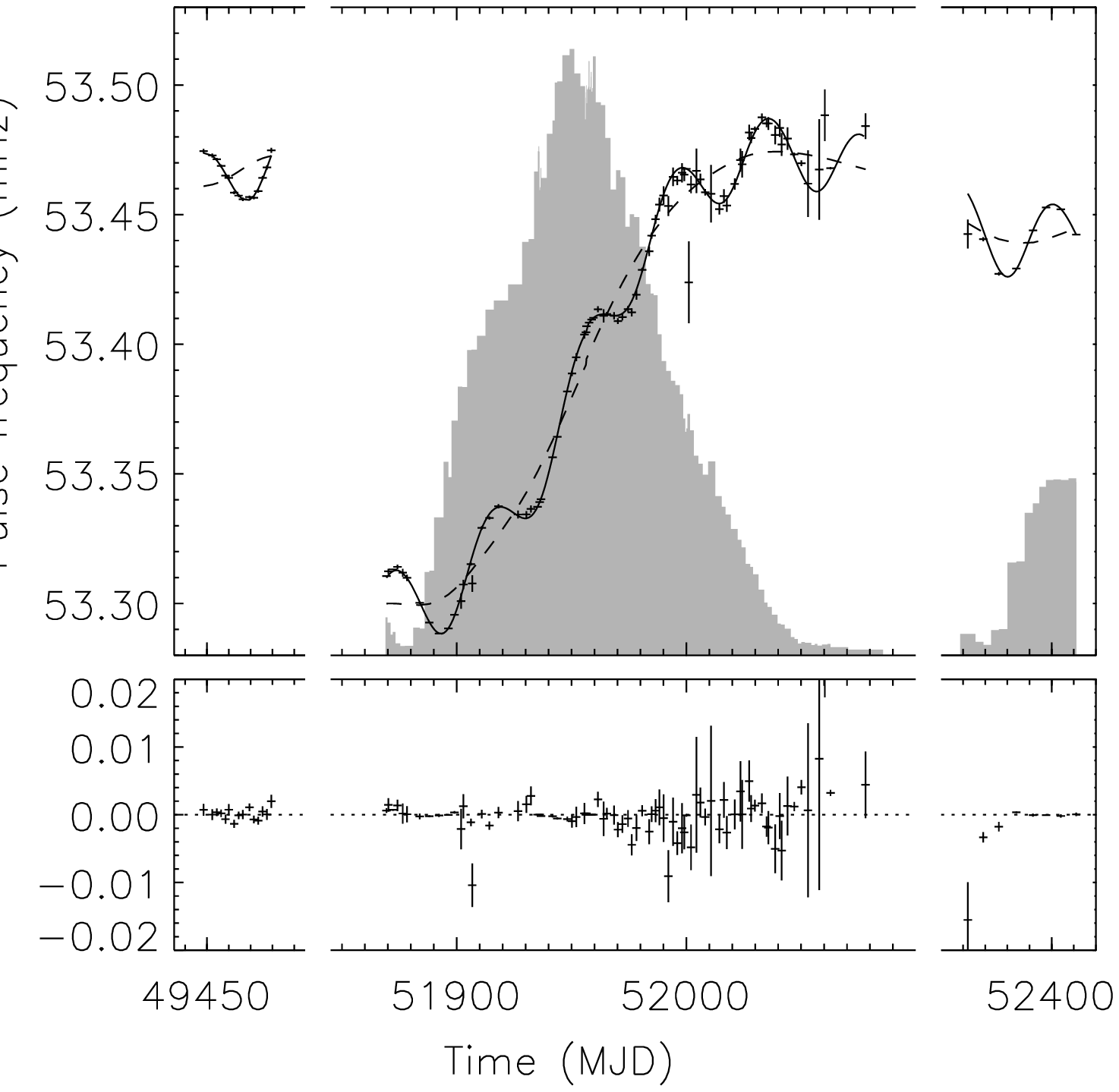}}
 \figcaption[f2.eps]{ Barycenter-corrected pulse frequency for
KS~1947+300 throughout the 1994 April outburst as measured by BATSE
\citep{chak95}, and the 2000--01 and 2002 outbursts as seen by {\it
RXTE}. The upper panel shows the frequency measurements and the
best-fit model (solid line; see \S\ref{freq}).  The sum of model
components representing the intrinsic neutron star spin frequency is
shown as a dashed line.  The lower panel shows the fit residuals.
Error bars represent 1$\sigma$ uncertainties.  The grey histogram
shows the PCA/HEXTE flux measurements.
 \label{period} }
\bigskip
\noindent
variation is superposed on this
increase; we naturally attribute it to Doppler shifts from orbital
motion in a binary system.

\subsection{Frequency Model}
\label{fmodel}

We fit the frequency measurements with a simple model comprising the
effects of neutron star spin including the spin-up due to accretion
torques and the apparent changes due to orbital Doppler shifts:
\begin{equation}
f(t) = f_{spin}(t) -
   \frac{2\pi f_{0j} a_{\rm x} \sin i}{P_{\rm orb}}(\cos l+g \sin 2l+h \cos 2l)
 \label{orbeq}
\end{equation}
where $f_{spin}(t)$ is the time-dependent stellar spin frequency,
$f_{0j}$ is an initial spin frequency 
for each of the three observation epochs (BATSE, \xte\/ 2000-01, and
\xte\/ 2002) denoted symbolically by $j$,
$a_{\rm x} \sin i$ is the projected orbital semimajor axis in units of
light travel time, and $P_{\rm orb}$ is the orbital period.  The
coefficients $g\ (=e \sin \omega)$ and $h\ (=e \cos \omega)$ are to be
interpreted as functions of the eccentricity $e$ and the longitude of
periastron $\omega$.  Finally, $l=2\pi(t-T_{\pi/2})/P_{\rm orb}+\pi/2$ is
the mean longitude, with $T_{\pi/2}$ the epoch at which the mean longitude
$l=\pi/2$.  For a circular orbit, $T_{\pi/2}$ is the epoch of superior
conjunction of the neutron star.  The right-most term in parentheses of
Eqn. 2 represents the orbital Doppler shifts to first order in $e$; given
the magnitudes of the uncertainties of our measurements, this should be an
adequate approximation as long as $e\la0.2$.

The neutron star spin frequency may be affected by several physical
mechanisms.  Among these, accretion torques are likely to be most
important.  As a first approximation, we take the accretion torque to
be proportional to a power of the observed X-ray flux, i.e.
$\tau_{\rm acc} \propto F_{\rm X}^\alpha$.  This prescription is a
limited generalization of models like those of \citet{gl79a,gl79b} in
which the torque is expected to be proportional to the accretion rate
and, in turn, the flux raised to the power $6/7$.

To obtain a simple model of the X-ray flux $F_{\rm X}$, we fitted a
composite light curve consisting of daily averages of PCA/HEXTE 2-60
keV fluxes and ASM 1.5-12 keV intensities (renormalized to 2-60 keV
fluxes) with the sum of a set of Gaussian components. The 2000--2001
outburst was modelled by the sum of 5 components, while the time
interval of the 2002 observations was modelled by a single component.
The parameter values defining the model are listed in Table
\ref{gaussflux}.  The individual components that describe the
2000--2001 outburst, as well as the total model, are plotted with the
measured flux values in Fig. \ref{asmlc}, lower panel.  Over the
2000--2001 outburst, the root-mean-square difference between the
measured fluxes and the model comprising five Gaussian components
(nos. 2-6 of Table 1) was $0.32 \times 10^{-9}\ \epcs$.
We also adopted an additional Gaussian component to represent the flux
evolution during the 1994 BATSE observations.  For that outburst,
which lasted around 30~d, the source reached a maximum pulsed
20--60~keV flux of around $4.5\times10^{-10}\ \epcs$ on MJD~49465
\cite[]{chak95}.  Broadband PCA/HEXTE spectra from the 2000--2001
outburst indicate that this likely corresponds to a 2--60~keV flux of
around $1.9\times10^{-9}\ \epcs$ for a pulse fraction of 50\% (see
\S\ref{ptime}). Thus, we included a Gaussian component centered on
MJD~49465 with a $1 \sigma$ width of 10~d and peak flux of
$1.9\times10^{-9}\ \epcs$.

We first tried to fit the measured frequencies with a simple model
based on Eqn. 2 in which the time integral of $\tau_{\rm acc}$ gives
the change in intrinsic spin frequency.  The best fit was poor
($\chi^2=2715$ for 101 degrees of freedom).  There are a number of
possible reasons for this, including (1) the spacing of the
observations does not allow us to construct a flux model that reliably
follows flux variations on short time scales ($\lesssim 10$ days),
(2) the accretion rate may not be precisely
proportional to the X-ray flux, and (3) the accretion torque in this
simple model vanishes when $F_X=0$. The latter property contradicts
the observation of a net spin down during the $\approx 6$ years between
the BATSE and \xte\/ observations while the source was in quiescence,
i.e. while $F_X\approx0$. In order to allow the net model torque to be
negative when $F_X=0$, we have augmented the spin model with terms
linear in time separately for each of the three epochs.
The revised model of the intrinsic spin
frequency is thus:
\begin{equation}
f_{spin}(t) = f_{0j} + b_j(t - t_{0j}) +
 \beta\int_{t_{0j}}^t F_{\rm X}^\alpha(u) du + \Delta
 f_{\rm gl}(t) \label{torqeq}
\end{equation}
where $f_{0j}$ and $t_{0j}$ are the frequency and time, respectively,
of the first frequency measurement for epoch $j$ and $F_{\rm X}(t)$ is
the 2-60 keV flux in units of $10^{-9}\ \epcs$.
The adjustable parameters $f_{0j}$ and $b_j$ are set independently for
each epoch while the power law index $\alpha$ and torque coefficient
$\beta$, along with the orbital parameters, are adjustable
parameters that apply globally to all three epochs.  Further 
analysis (see \S\ref{ptime}) indicated that there was a rapid change in the
intrinsic pulse
\centerline{\epsfxsize=8.5cm\epsfbox{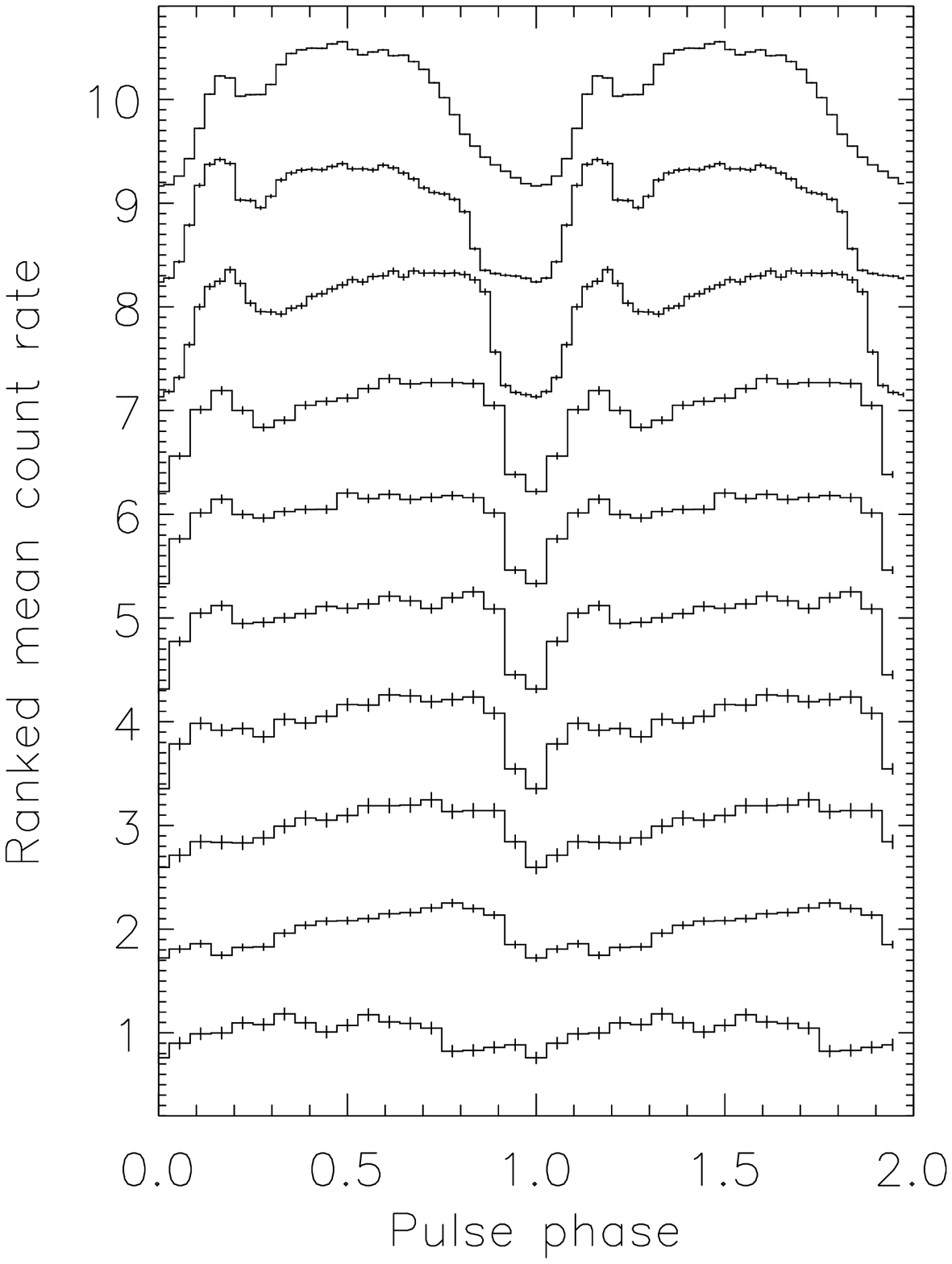}}
\figcaption{ Representative pulse profiles from \src\ ranked by mean
count rate. From top to bottom, the profiles are for the observations
of 2001 Feb 10, Apr 5, May 6, May 12, May 17, May 18, May 21, May 27,
2000 Nov 28, and 2001 Jun 18
(MJD 51950, 52004, 52035, 52041, 52046, 52047, 52050, 52056, 51876 and
52078),
respectively. The profiles have been
rescaled by the mean rate and shifted in phase so that the minimum
falls at phase 0.0. Each profile is plotted twice and shifted
vertically for clarity.
\label{prof} }
\bigskip
\noindent
frequency around $t_{\rm gl} \simeq {\rm MJD}\ 51956.41$.  We therefore 
included the term $\Delta f_{\rm gl}(t)$ which was assigned a
value of zero for all measurements made prior to $t_{\rm gl}$ and a
constant value of $1.83\ \mu$Hz for all measurements made afterwards.  The
magnitude of this frequency change was determined
through a pulse arrival time analysis which is described below, and so was
not a free parameter in the frequency model.

The model can be formulated so that it depends linearly on all the
adjustable parameters except for $\alpha$ and $P_{\rm orb}$.  Thus, we
searched a grid of points in $\alpha$-$P_{\rm orb}$ space to determine
the best-fitting parameter combination.  When obtaining the best fit
for each $\alpha$-$P_{\rm orb}$ pair of values, all other parameters
were allowed to vary in a linear $\chi^2$ fitting process.  The
values of the parameters for the best-fit model are listed in Table
\ref{orbit}.  The model 
fits the data much better than our
original simple model described above, although there are still significant
residuals (Fig.  \ref{period}).  This is reflected in the reduced
$\chi^2$ value of $3.35$.  The fit results unambiguously
yield the number of orbital cycles from the epoch of the BATSE
observations to that of the \xte\/ observations, and thus an accurate
orbital period of 40.415~d.  We also note that the values of the coefficients
$b_j$, i.e., the coefficients of the terms representing the values of
$\dot{f}$ at zero flux for each epoch, are negative as expected from
the spin down inferred during quiescence.

We first estimate the uncertainty in each fitted parameter of the
frequency model from the range of values of that parameter that occur
within the region defined by $\chi^2 = \chi^2_{min} + \chi^2_{\nu}$
that we obtain when all parameters, except $\alpha$, are allowed to
vary.  Here $\chi^2_{\nu} = 3.35$ is the value of $\chi^2$ per degree
of freedom for the best fit.  This $\chi^2$ contour is that which
yields $1 \sigma$ errors when the uncertainties in the individual
frequency measurements are scaled to give $\chi^2_{\nu} = 1$, assuming
correlations in the errors of different measurements are negligible.
These estimates are given in column 4 of Table 2. These uncertainties
do not reflect errors that could occur because (1) there could be
correlations in the errors in the frequency measurements, (2) the flux
model may not be precise, or (3) the model may not be formulated
properly.  We have attempted to crudely estimate the effects of the
second possibility and the issue of possible systematic errors in
general by varying the parameters describing the flux model one at a
time while fixing all other flux model parameters at the values given
in Table 1 and then redoing the fit for each new set of values.  In
this exercise we did not limit the range of values of the modified
parameter to those values that give a good fit to the flux data.  The
magnitudes of the variation of each derived frequency model parameter
in this exercise are given in column 5 of Table 2.

The fitted eccentricity of 0.033 indicates that the orbit is almost
circular.  Even so, a circular orbit fit ($e=0$) results in a
significantly poorer fit ($\chi^2_\nu = 4.57$).  The formal
probability that the fit would yield $e \geq 0.033$ when the true
orbit is perfectly circular is $\sim 1.1\times10^{-7}$; this is
equivalent to a detection of non-zero $e$ with $5\sigma$ confidence
(cf. Lucy \& Sweeney 1971). However, we note that the systematic
uncertainty estimated in the exercise described above suggests that
the significance of the measured eccentricity is closer to $\sim
2.5\sigma$.

\subsection{Pulse Arrival Time Model}
\label{ptime}

The pulse frequency fit takes advantage of the long baseline
spanned by the observations, and yields an accurate determination of
$P_{\rm orb}$.  However, that analysis did not take advantage of the
relative phases of the pulsations from observation to observation.  We
therefore undertook a pulse arrival time analysis of the PCA data
obtained in the first epoch of \xte\ observations.

We first constructed a pulse profile for each observation by folding data
in 0.25-s time bins using the pulse period predicted
by a preliminary but reasonably accurate frequency model that included
both intrinsic frequency changes and orbital Doppler shifts.  Profiles for
selected observations are shown in Fig. \ref{prof}.  The pulse shape
evolved as the flux increased and decreased.  At lowest flux the profile
exhibited a single peak, with a narrow dip spanning 0.2 in phase forming
the primary minimum.  As the flux increased, a shallower secondary dip
around 0.25 later in phase developed, so that the profile exhibited one
narrow and one broad peak.  The primary minimum also became progressively
broader and asymmetric, with a more gradual ingress and egress.  As the
flux decreased following the peak, the same systematic variation in pulse
profile was retraced in the opposite direction.

For the determination of pulse arrival times, we folded data from
512-s segments.  We defined the pulse arrival
time as the epoch of
maximum flux, which was determined by fitting a sinusoid to the pulse
profile. The errors in the arrival times were estimated by varying the
epoch of maximum flux in the fit over a grid to determine the interval
in which $\Delta\chi^2\leq1$.

To make effective use of these arrival times, one must determine the
number of pulsation cycles between each measured pulse arrival time.
Thus, we used the preliminary frequency model noted above to estimate
cycle counts.  When observations were separated by longer than $\sim
2$~d it was not possible to unambiguously determine the cycle count
using the frequency data alone.  Instead, we performed a joint fit of
both the frequency and arrival time data. For each long data gap, we
fitted both frequencies and arrival times before and after the
gap for times as far forward and backward as possible without
introducing additional cycle-count ambiguities. Using this approach,
we were able to determine the cycle counts from MJD~51914 to
MJD~52062.

When the expression on the right in Eqn. 3 is substituted into
Eqn. \ref{orbeq} and integrated, one obtains an expresssion for
the pulse phase.  Pulse arrival times may then be taken to be the
times when the phase is an integer number of cycles.  A constant of
integration appears in going from frequency to phase; it may be taken
as the arrival time of a given pulse.

We then fit the arrival times within the range over which we can
unambiguously phase-connect the observations, as well as the frequency
data before and after.  The best-fit model did not fit the data well,
as indicated by the large value of reduced $\chi^2$.  One major reason
for the lack of agreement between the model and the measured arrival
times was an instance, around MJD~51956.4, where the pulse frequency
changed rather rapidly (see Fig. \ref{glitchfig}).  The event is
apparent in both the frequency measurements and in the arrival time
(or, equivalently, pulse phase) residuals from a fit calculated using
only pulse arrival times prior to the event.  Examination of pulse
profiles revealed no significant pulse shape change around the time of
the event.

The residuals were consistent with a stepwise change of frequency of
$\Delta f\approx+$\delf\ occurring around \tglitch\ (such a change is
included in the frequency fit; see \S\ref{fmodel}).
In order to obtain
limits on the magnitude and duration of this event, we fit a
model to the arrival time data that included a step in frequency
$\Delta f_{\rm gl}$ that occurred via a constant frequency derivative over
a time interval $\Delta t_{\rm gl}$ centered at time $t_{\rm gl}$.  These
parameters, as well as low-order polynomial terms in time, were
allowed to vary in the fitting process while the orbital parameters
and the torque coefficient were fixed at the values determined from
the best-fit frequency model.  When this fit was performed on the
arrival time data from the 7.8 day time interval MJD 51952.03
through 51959.83, we obtain $\Delta f_{\rm gl} = 2.01 \pm 0.04\ \mu$Hz,
$t_{\rm gl} =$ MJD $51956.393 \pm 0.01$, and $\Delta t_{\rm gl} < 0.4$ days,
where the errors have been scaled by the square root of $\chi^2_\nu =
0.69$.  When this fit was performed on the arrival time data from a
shorter time interval, i.e., MJD~51955.6 through MJD 51959.1, the best
fit is obtained for $\Delta f_{\rm gl} = 1.83 \pm 0.09\ \mu$Hz, $t_{\rm gl} =$
MJD $51956.416 \pm 0.014$, and $\Delta t_{\rm gl} < 0.5$ days, again where
the errors have been scaled by the square root of $\chi^2_\nu =
0.67$. The best-fit values of the parameters which describe the glitch
\centerline{\epsfxsize=8.5cm\epsfbox{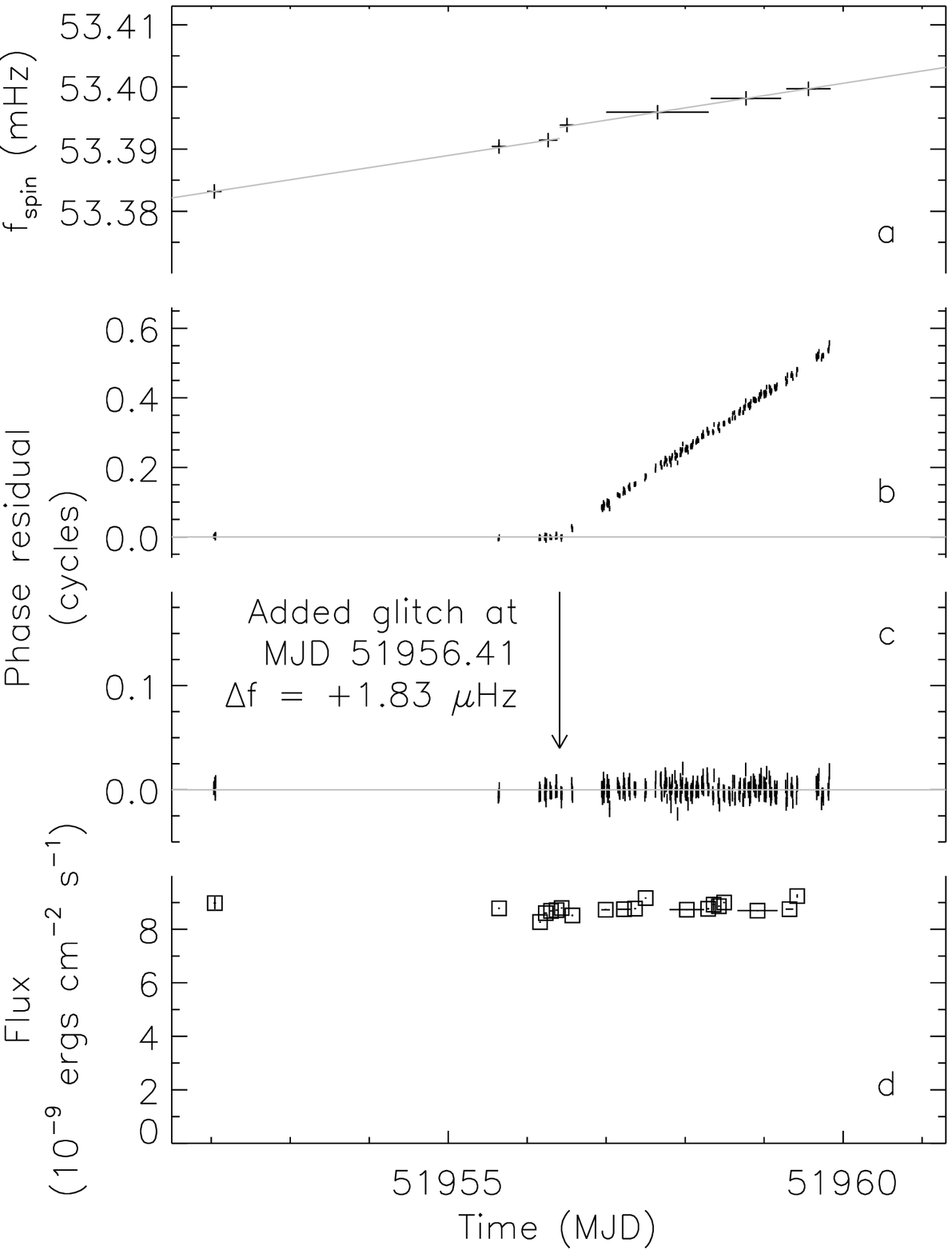}}
  \figcaption[glitch.eps]{Pulse frequencies, phase residuals, and
X-ray intensities around the time of the glitch.  ({\it a}) Measured
pulse frequency corrected for the binary orbit with error bars
corresponding to $3\sigma$ confidence intervals.  ({\it b}) Phase
residuals calculated from a best-fit model that does not include a
component representing the glitch.  ({\it c}) Phase residuals from a
full model that does include a component representing the glitch. Note
the difference in scale between panels {\it b} and {\it c}.  ({\it d})
The 2-60 keV band source flux.
 \label{glitchfig} }
\bigskip
\noindent
do not appear to depend, within limits, on the length of the data set.
The full sets of model parameters of both fits may be found in Table
3. We note that the parameters listed in Table 3 are 
consistent with an instantaneous change in frequency.

\section{DISCUSSION}

We measured the flux and pulse frequency evolution of the X-ray
pulsar \src\ through a large outburst in 2000--01, as well as in a
smaller outburst in 2002.  We fitted the measured pulse frequencies and
arrival times to derive orbital parameters, and to determine the
temporal variation of the intrinsic spin frequency of the neutron
star.  We found that the pulsar revolves in a nearly circular orbit
($e=\ecc$) with orbital period \porb, and projected semi-major axis \axsini.
This orbital period is significantly shorter than that derived from
analysis of the pre-outburst ASM light curve, $41.7 \pm 0.1$ days
\cite[]{levine00}.  Although it is clear that the value based on
pulse timing is correct, the basis of the discrepancy is not
clear.  We note that the X-ray light curves of other Be-star/neutron star
binaries sometimes do contain quasiperiodic variations that are
asynchronous with the orbits (e.g., GS~0834$-$430, Wilson et al. 1997;
XTE~J1946+274, Wilson et al. 2003).

If we assume that the neutron star mass $M_{\rm X}=1.4
M_\sun$, then the X-ray mass function of $f_{\rm X}=$\fxm\ implies a
companion mass of at least 3.4 $M_\sun$ (for an inclination
$i \le 90\arcdeg$).  We do not find eclipses in either the ASM or the PCA
light curves, and can confidently conclude that the system is not
eclipsing, thus ruling out inclinations $\ga 85\arcdeg$.  Companion
masses $M_{\rm c} =$ 5, 10, or 20 $M_\sun$, imply inclinations of
$57\arcdeg$, $38\arcdeg$, or $28\arcdeg$ respectively.  This range of
masses and inclinations is compatible with the B0Ve spectral type of
the optical counterpart \cite[]{neg02}.

The wide orbital separation and low eccentricity place \src\ within
the small class of neutron stars with high-mass binary companions
which almost certainly did not receive large ``kicks'' during the
supernovae which led to their formation \cite[]{pfahl02}.
The optical counterpart appears to be a Be star that is not highly
evolved.  Thus its radius is most likely much smaller than the radius
of its Roche lobe, i.e., than 104, 141, or 191 lt-s for $M_{\rm X} =$
1.4 $M_\sun$ and $M_{\rm c} =$ 5, 10, or 20 $M_\sun$, respectively.
Furthermore, the wide separation and relatively small size of the
companion suggest that it is unlikely that tidal effects have
circularized the orbit since the supernova event wherein the neutron
star was formed.  If the pre-supernova orbit was circular, the
measured eccentricity $e \sim 0.03$ suggests that the mass ejected in
the supernova must have been rather small, i.e., $\Delta M / M_{\rm final}
< 0.03$, where $M_{\rm final}$, the post-supernova system mass, could be
as large as $20 M_\sun$.  The low eccentricity of this system clearly
presents a number of difficulties.

An outburst of \src\ was seen with BATSE in 1994, and no other
outburst was reported until that of 2000--2001.  BATSE operated
continuously until 2000 May, while the \xte\ ASM has operated more or
less continuously since early 1996.  BATSE would most likely have
detected any pulsed flux greater than $\approx 20$~mCrab in the 30--75
keV band (equivalent to a flux $2 \times 10^{-10}\ \epcs$) that
was present for more than a few days. The ASM would most likely have
detected any outburst longer than $\approx 5$~d that reached a strength
of $\approx 1$~ASM~count~s$^{-1}$ (equivalent to $13$~mCrab or to a
2-12 keV band flux $3 \times 10^{-10}\ \epcs$).
Thus, as was also discussed by \cite{neg02}, we can rule out the
occurrence of a moderate to strong outburst in the intervening time,
although it is possible that weak outbursts could have been missed.

From the end of the outburst in 1994 to the beginning of the outburst in
late 2000, the pulsar frequency decreased from 53.47 to 53.30~mHz.  The
average frequency derivative during this interval was $\dot{f} = -0.8
\times 10^{-12}$ Hz s$^{-1}$. Spin-down in quiescence likely arises from
the dominance of magnetic torques contributed by field lines which thread
the accretion disk outside the corotation radius
\cite[e.g.][]{gl79a,gl79b}. Alternatively, negative torques may arise from
the ``propeller effect'' \cite[]{is75}. We note that the magnitude of the
quiescent spin-down was comparable to that measured for at least one other
pulsar, A~0535+26, of $-(2.2\pm0.6)\times10^{-13}\ {\rm Hz\,s^{-1}}$
\cite[]{finger94}.
While the values of $b_j$ for each epoch (see Table 2) are comparable to,
or significantly larger than the quiescent spin-down, we caution that it
is not correct to interpret these parameters independently from the torque
model.  The $b_j$ are coefficients of terms we used to augment the
accretion torque integral in equation \ref{torqeq}, so as to more
accurately model the intrinsic spin changes of the neutron star.  The
integral alone would give zero torque only at zero flux, whereas the
various negative fitted values of the $b_j$ indicate that zero torque
corresponds to non-zero flux at a different level for each epoch.

The 2--60~keV flux of \src\ peaked at $9 \times 10^{-9}\ \epcs$ in
2001 February.  This flux implies an X-ray luminosity of $1.1 \times
10^{38}\ \eps$ for a source distance of 10~kpc.  Given this accretion
luminosity, a neutron star of mass $M_{\rm X} = 1.4\ M_\sun$, radius
$R = 10$~km, magnetic moment $\mu = 1 \times 10^{30}$ G cm$^3$
(equivalent to an equatorial surface magnetic field strength of
$10^{12}$~G), and moment of inertia $I = 1 \times 10^{45}$ g cm$^2$, a
standard theoretical treatment \cite[Eqn. 15 of][]{gl79b,jsrp76}
predicts $\dot{f} = 1.5 \times 10^{-11}$ Hz s$^{-1}$ whereas the
intrinsic frequency derivative around the time of maximum flux
obtained from our pulse timing results is $\dot{f} \simeq 2.3 \times
10^{-11}$ Hz s$^{-1}$.  These values are consistent given the simple
nature of the theoretical estimate, the possibility of deviations from
the assumed neutron star parameters, and the uncertainty in the
distance to the system.

We found evidence for a particularly rapid variation of the pulse
frequency around \tglitch.  The phase residuals around this episode
were minimized by including a change in frequency of $\Delta
f\approx+$\delf\ that occurred at a uniform rate over a time interval
$\Delta t \lesssim 10$~hr.  The frequency change corresponds to a
fractional change $\Delta f/f \approx$\ \delffrac.  If the frequency
change occurred at a constant rate over a 10 hr interval, then
the intrinsic neutron star pulse frequency derivative must have reached at
least $4.8\times10^{-11}\ {\rm Hz\,s^{-1}}$, more than twice as large as
the maximum inferred from the torque model, \maxfdot.  We note that this
event occurred close to the time of peak flux and, therefore,
when the accretion torque was near its maximum value.  While it is
possible that the abrupt increase in frequency arose from a brief episode
of enhanced accretion, we found no evidence for the increase in flux which
might be expected as an immediate consequence (see Fig.  \ref{glitchfig}).
A transient flux increase could have been missed during the $\approx4$~hr
gap (MJD~51956.39 to 51956.56) between the observations closest to the
estimated time of the glitch. However, the increase in X-ray luminosity
required to explain the magnitude of the glitch would be a factor of
$\approx2$, sufficient to exceed the Eddingon limit for a canonical
neutron star at 10~kpc distance.  Furthermore, four ASM dwells at
MJD~51956.51 indicate a flux of $7.3\pm0.6\ \cts$, consistent with the level
before and after the gap.

This event may have arisen from a sudden change in the spin
frequency of the neutron star, analogous to the ``glitches''
observed in other types of neutron stars.
However, there are also some significant differences between the glitch in
\src\ and those in radio pulsars and AXPs.  First, in contrast to radio pulsar
glitches, which frequently exhibit a change in the spin-down rate
$\dot{f}$ following a glitch, we measured consistent values of $\dot{f}$
before and after \tglitch.  Second, the fractional change in frequency of
\delffrac\ in the \src\ glitch is around an order of magnitude larger than
the typical value for ``giant'' glitches in radio pulsars
\cite[]{glitch03}. Finally, what is more puzzling is that the frequency
change in \src\ was in the same sense as the mean spin-up around that
time, i.e.  the glitch resulted in a brief {\it acceleration} of the
spin-up rate by at least a factor of 2 to 3. Since the neutron star was
spinning up at the time of the glitch, one might expect that the core
would be rotating more slowly than the crust, so that a glitch would cause
a sudden decrease in the pulse frequency.  However, a long history of
spin up and spin down episodes may have left the core spinning faster than
the crust at the time of the glitch.  The BATSE observations indeed show
that the pulse frequency in 1994 was higher than the pulse frequency at
the time of the glitch.

This is the first reported glitch in an accretion-powered pulsar.  If
such events occur in other accreting neutron stars, why have they not been
detected previously?
The main reason may be that such events are simply difficult to detect;
indeed, there are a number of conditions which are required for the
detection of glitches in either accreting or rotation-powered neutron
stars.  The pulsars must be monitored over a period of days, with
observations being done at least several times per day.  This is often
more difficult to accomplish from space-based X-ray satellites than
from ground-based radio observatories.  Also, the extraction of pulse
frequencies from previous X-ray observations of pulsars have sometimes
been done using 1 day or longer integration time intervals, which
makes it difficult to detect $\mu$Hz frequency changes occurring on
time scales of hours or less.  Sudden frequency changes could be lost
among the other frequency changes from torque variations which
typically take place on longer time scales.

We can estimate the sensitivity for the detection of a sudden
frequency change from a set of observations.  For this purpose we
assume that two estimates of the pulse frequency are made, one before
and one after the time of a possible sudden change.  For simplicity,
we further assume that each frequency estimate is made from a set of
$N$ equally-spaced observations made
over an interval of length $T$, that each observation yields a phase
estimate with $1\sigma$ uncertainty $\sigma_\phi$, and that there are no
pulse counting ambiguities left unresolved.  In this case the $1\sigma$
uncertainty in the difference of the two frequency estimates is given by
\begin{equation}
\sigma_f \sim a \left( \frac{\sigma_\phi}{0.1\ {\rm cycle}} \right)
   \left( \frac{T}{1\ {\rm d}} \right)^{-1} N^{-1/2}\ \mu{\rm Hz}
 \label{uncert}
\end{equation}
where $a$ is a factor of order unity.  The length of the time interval
$T$ over which the frequency estimates are to be made is limited by
the frequency changes generally ascribed to accretion torque
variability.  Equation \ref{uncert} indicates that the detection of a $\sim 1\
\mu$Hz glitch requires, for example, $N\sim10$ observations over the
course of a few days with each observation yielding a modest accuracy
phase measurement.

The BATSE instrument on the Compton Gamma-Ray Observatory monitored
many accretion-powered pulsars over the course of 9 years with a
sensitivity of $\sim 1\ \mu$Hz per day in favorable cases (Bildsten et
al. 1997).  Frequencies were reported for every day or every few days
that pulsations were detected (Chakrabarty et al. 1997a, Finger et
al. 1999, Koh et al. 1997, Chakrabarty et al. 1997b).  Though the
BATSE results have not led to the report of any glitch, it is possible
that some of the monitored pulsars did exhibit glitches that were not
easily noticed because of the large frequency variations on time
scales of days and weeks.

A second possibility is that such events may also be quite rare.
In this regard, we note that \src\ was accreting near the Eddington
limit when this glitch occurred.  If glitches in accretion-powered
pulsars only occur under such circumstances, then the opportunities
for their detection would be quite limited, since the times when
Galactic pulsars are observed to be accreting near the Eddington limit
are relatively infrequent.
Clearly, new analyses of archival data and additional observations are
called for in order to find other examples of glitches and to determine
how often such events occur in accreting neutron stars.

\acknowledgments

This research has made use of data obtained through the High Energy
Astrophysics Science Archive Research Center Online Service, provided
by the NASA/Goddard Space Flight Center.  We thank D. Chakrabarty for
providing pulse frequencies from the 1994 BATSE observations and for
helpful comments on the manuscript, and S. Rappaport for helpful
comments.  This work was supported in part by NASA Contract NAS5-30612
and by the NASA Long Term Space Astrophysics program under grant NAG
5-9184.


\clearpage

\begin{deluxetable}{lcccc}
\tablecaption{Gaussian Components of the X-ray Light Curve
   \label{gaussflux}}
\tablewidth{0pt}
\tablehead{ & \colhead{Component} & & & \colhead{Peak Intensity} \\
 \colhead{Data set} & \colhead{number} & \colhead{Center (MJD)}
  & \colhead{Width\tablenotemark{a} (d)} & \colhead{($10^{-9}\ \epcs$)} }
\startdata
BATSE &   1 & 49465.0 & 10.0 & 1.92 \\
\xte\/  & 2 & 51855.0 & 10.0 & 1.92 \\
        & 3 & 51904.2 & 10.0 & 2.72 \\
        & 4 & 51945.4 & 23.6 & 6.43 \\
        & 5 & 51974.9 & 32.1 & 4.12 \\
        & 6 & 52020.5 &  7.4 & 0.449 \\
        & 7 & 52405.0 & 17.0 & 2.88 \\
\enddata
\tablenotetext{a}{Standard deviation}
\end{deluxetable}


\begin{deluxetable}{lcccc}
\tablecaption{Frequency model parameters for KS~1947+30\label{orbit}}
\tablewidth{0pt}
\tablehead{
  & & & & \colhead{Uncertainty due} \\
  \colhead{Parameter} &  \colhead{Epoch}
 & \colhead{Fit Value}  & \colhead{Uncertainty\tablenotemark{a}} 
 & \colhead{to Flux Model\tablenotemark{b}}  }
\startdata
$t_{01}$ (MJD)                 & BATSE & $49448.5$ & \nodata & \nodata \\
$f_{01}$ (mHz)                 & BATSE & 53.4610  & 0.0015 & 0.0010 \\
$b_1$ ($10^{-12}$~Hz\ s$^{-1}$) & BATSE & -1.9 & 0.9 & 6.2 \\
$t_{02}$ (MJD)                 & \xte\ 1 & $51869.397$ & \nodata & \nodata \\
$f_{02}$ (mHz)                 & \xte\ 1 & 53.3000 & 0.0003 & 0.0005 \\
$b_2$ ($10^{-12}$~Hz\ s$^{-1}$) & \xte\ 1 & -3.4 & 0.2 & 0.6 \\
$t_{03}$ (MJD)                 & \xte\ 2 & $52362.115$ & \nodata & \nodata \\
$f_{03}$ (mHz)                 & \xte\ 2 & 53.4448 & 0.0008 & 0.0036 \\
$b_3$ ($10^{-12}$~Hz\ s$^{-1}$) & \xte\ 2 & -7.2 & 0.3 & 3.5 \\
$t_{\rm gl}$ (MJD)              & All & 51956.416 & \nodata & \nodata \\
$\Delta f_{\rm gl}$ (Hz)        & All & $1.83 \times 10^{-6}$ & \nodata
  & \nodata \\
$\alpha$                    & All & 0.75 & 0.02 & \nodata \\
$\beta \tablenotemark{c}$   & All & $4.35 \times 10^{-7}$ &
   $0.07 \times 10^{-7}$ & $0.21 \times 10^{-7}$ \\
$P_{\rm orb}$ (d)           & All & 40.415 & 0.007 & 0.010 \\
$a_{\rm x} \sin i$ (lt-s) & All & 137.4 & 1.2 & 3.0 \\
$T_{\pi/2}$ (MJD)           & All & 51985.31 & 0.07 & 0.06 \\
$e$                         & All & 0.034 & 0.007 & 0.013 \\
$\omega$ ($\arcdeg$) & All & 33 & 3 & \nodata \\
$f_{\rm X}(M)$ ($M_\sun$)   & All & 1.71 & 0.04 & \nodata\\
\tableline
$\chi^2/N$(DoF) & \multicolumn{4}{c}{$329/98$}

 \enddata

\tablenotetext{a}{The range of variation of the parameter values that
occur within the $\chi^2 = \chi^2_{min} + \chi^2_{min}/$N(DoF) region
that applies when all other frequency model parameters, except for
$\alpha$, are allowed to vary.}

\tablenotetext{b}{The range of variation of the parameter for the best
fit when the parameters describing the flux model (Table 1) are varied
one at a time (see text).}

\tablenotetext{c}{Torque coefficient in units of 
   Hz day$^{-1}$} 

\end{deluxetable}

\begin{deluxetable}{lcc}
\tablecaption{Pulse Arrival Time Models Around the Glitch\tablenotemark{a}
   \label{patmodel}}
\tablewidth{0pt}
\tablehead{
  \colhead{Parameter} &  \colhead{Value\tablenotemark{b} -- Fit 1} &
  \colhead{Value\tablenotemark{b} -- Fit 2}  }
\startdata
Time interval (MJD)         & 51952.03--51959.83 & 51955.6--51959.1 \\
$t_0$ (MJD)                 & $51869.397$ & $51869.397$ \\
$f_0$ (mHz)                 & $53.29791 \pm 0.00027$ & $53.2976 \pm 0.0014$ \\
$b$ ($10^{-12}$~Hz\ s$^{-1}$) & $-15.6 \pm 0.2$ & $-15.4 \pm 1.0$ \\
$t_{\rm gl}$ (MJD)              & $51956.393 \pm 0.010$ & $51956.416 \pm 0.014$ \\
$\Delta t_{\rm gl}$ (days)      & $< 0.4$ & $< 0.5$ \\
$\Delta f_{\rm gl}$ ($\mu$Hz)   & $2.01 \pm 0.04$ & $1.83 \pm 0.09$ \\
\tableline
$\chi^2/$N(DoF)             & $154.0/224$ & $121.7/183$ \\
\enddata

\tablenotetext{a}{The following parameters were held fixed at the
indicated values determined from the frequency model fitting: $P_{\rm
orb} = 40.415$ d; $a_{\rm x} \sin i = 137.3$ (lt-s); $T_{\pi/2} =
(MJD)~51985.31$; $g = 0.018928$; $h = 0.027032$; $\alpha = 0.75$;
$\beta = 4.35 \times 10^{-7}$ Hz day$^{-1}$.
The values of $g$ and $h$ correspond to an
eccentricity $e = \sqrt(g^2 + h^2) = 0.033$.  The uninteresting
parameter $\phi_0$ is not listed here.}

\tablenotetext{b}{Parameters without uncertainties were held fixed at
the given values.  The indicated uncertainties correspond to $1
\sigma$ confidence levels obtained from evaluation of $\Delta \chi^2 =
\chi^2_\nu$.  Upper limits correspond to $2 \sigma$ confidence limits.}

\end{deluxetable}

\end{document}